\documentclass{acm_proc_article-sp}
\usepackage[utf8]{inputenc}
\inputencoding{utf8}

\begin{document}

\title{Inferring gender of a Twitter user using celebrities it follows}

\numberofauthors{1} %  in this sample file, there are a *total*
\author{
% 1st. author
\alignauthor
Puneet Singh Ludu
\\
\affaddr{State University of New York}\\
\affaddr{Buffalo, New York}\\
\email{pludu@buffalo.edu}
}

% There's nothing stopping you putting the seventh, eighth, etc.
% author on the opening page (as the 'third row') but we ask,
% for aesthetic reasons that you place these 'additional authors'
% in the \additional authors block, viz.

% Just remember to make sure that the TOTAL number of authors
% is the number that will appear on the first page PLUS the
% number that will appear in the \additionalauthors section.

\maketitle
\begin{abstract}
This paper addresses the task of user gender classification in
social media, with an application to Twitter. The approach automatically predicts gender by leveraging observable
information such as the tweet behavior, linguistic content of the user's Twitter feed and the celebrities followed by the user.
\\
This paper first evaluates linguistic content based features using LIWC dictionary and popular neighborhood features using Wikipedia and Freebase.
Then augments both features which yielded a significant increase in the accuracy for gender prediction. 
Results show that rich linguistic features combined with popular neighborhood prove valuables and promising for
additional user classification needs.
\end{abstract}

%\category{H.4}{Social Information retrieval, }{Miscellaneous}
%A category including the fourth, optional field follows...
%\category{D.2.8}{Software Engineering}{Metrics}[complexity measures, performance measures]


 % NOT required for Proceedings

\section{Introduction}

Twitter a micro-blogging site, have become an integral part of the life of millions of users. 
People use it to communicate with friends,family or acquaintances. With the increase in the number of users, demand for 
analytics on this data is also growing. Twitter does not store gender, age, ethnicity or interests of a user. 
These attributes of a user could be useful both for user experience as well as for consumption by brands in their social analytics.
\\
In this work we address the task of user gender classification, by leveraging observable
information such as the tweet behavior, linguistic content of the user's Twitter feed and the celebrities followed by the user.
\\
Main contributions of this work are following:
\begin{description}
\item[$\bullet$] It uses some new linguistic content based features, which were not used by state-of-the-art, particularly on Twitter data.
\item[$\bullet$] It describes completely new set of popular(celebrity) neighborhood based features such as \textit{age, gender and occupational area of the celebrity} which show promising results. 
\item[$\bullet$] It reports that using combination of both features can perform better than various state-of-the-art techniques, and especially extraction of better popular network based features demands further investigation.
\end{description}

The paper is organized as follows. Section 2 introduces relevant and related work on user profiling for social
media, Twitter user attribute detection and topic models for
Twitter. Section 3 describes the datasets we have used or created. Section 4 we explains methodology we followed to clean and process the datasets, while Section 5
explains in detail use of various features and the intuition behind their usage. Section 6 describes results produced in different configurations. Finally, Section 7 draws final conclusions and outline future work.

\section{Background and Related work}
Inferring attributes of social media users is a growing area of interest, there have been many recent attempts to predict various hidden attributes of a user. Several of the recent works were focused on predicting ethnicity (Rao et al., 2011; Pennacchiotti and Popescu, 2011), age (Schler et al., 2006; Rosenthal and McKeown, 2011; Nguyen et al., 2011; Al Zamal et al., 2012), gender (Rao et al., 2010; Burger et al., 2011; Liu and Ruths, 2013; Al Zamal et al., 2012), Interests ( Lim and Datta, 2013), personality (Argamon et al., 2005; Schwartz et al., 2013) etc.\\
Most of the datasets prepared in these approaches were hand annotated, cherry picked or attributes are identified by the user itself. Also many of the datasets were limited to very broad attribute categories, such as age below 23 as attribute young and above 25 as old. Rao et al. (2011) used linguistic features to predict ethnicity and gender of facebook users. Using a very limited training data, they tried to evaluate fine grained ethnicity classes of Nigeria. Social or Network based features were first used by Pennacchiotti and Popescu (2011); they tried to predict if a Twitter user is African-American or not.  \\
As preparing a good dataset is a big challenge, Schwartz et al. (2013) tried to explore this area and collected Facebook profiles labeled with personality type, gender, and age by administering a survey of users embedded in a personality test application.\\
Al Zamal et al. (2012) explored into social features by exploring homophily based features; such as using linguistic features of n-most-popular friends of a user on Twitter. Their final classifier obtains a \textbf{80.02\%} accuracy on binary gender classification (compared with the \textbf{85.67\%} accuracy this paper report below for same dataset).\\
Lim and Datta (2013) presented a novel approach for classifying user interests for e.g. Food, Politics, Charity etc. using celebrity(popular users) followed by a twitter users. \\

\section{Datasets}
In order to collect the datasets, it was necessary to decide on the two contrasting labels (for e.g ``male'' and ``female'') that would be used and identify users that could be reliably assigned one label or
the other. Even though we have used dataset which was already prepared by Zamal, still we had to manually inspect accounts collected
 to confirm that the labeled assignments were correct.

\subsection{Gender Dataset}
This paper have used the same dataset prepared by Al Zamal et al. (2012) for gender prediction. Zamal's paper used a technique proposed in Mislove et al. (2011), which used users who had their full first and last names on Twitter; also their first name was one of the top 100 most common names on record with US social security department for baby boys/girls born in the year 2011. 131 male and 134 female labeled users were collected out of 192 each, because some users closed their Twitter account or locked their timeline for public access.
\\
\subsection{Popular neighborhood Dataset}
We have also created a dataset of the popular neighborhood or so called celebrity users followed by these 265 users. Here, we define celebrities(popular users) as users with more than 10,000 followers or who have been verified by Twitter itself, as was defined by Lim and Datta (2013). Thus, out of 108,619 distinct users followed by 265 users, we were left with 5,546 popular users. Once we had the list of these celebrity users we extracted various features which we would discuss in section 5.

\section{Methodology}

The initial results of applying user tweets based classification 
did not seem very promising, a closer examination revealed
that the problem was in the quality of the Tweet text. Since the text
in this domain comes from Short messages, and the messages often contain various abbreviations, and are rife with
spelling errors. Thus, to deal with this, we implemented a process to auto-correct words that are greater than 3 letters in length as suggested by Prem Melville et al. (2013). This process leveraged a combination of Philips' metaphone algorithm and string-edit(TextBrew) distance for spell-correction; in addition to a general-purpose English dictionary, we used dictionary for Twitter short hands, 
for e.g. \textit{``idk''} would be mapped to \textit{``I don't know''}. As seen on Twitter many people use a lot of camel-casing especially while using hashtags,
we tried to clean such camel-cases, for e.g. \textit{``\#BestDayEver''} would be cleaned as \textit{``Best Day Ever''}.
After cleaning tweets we saw 12\% gain in accuracy only based on user tweets based classification. 

\begin{figure}[!htbp]
\centering
\includegraphics[width=80mm]{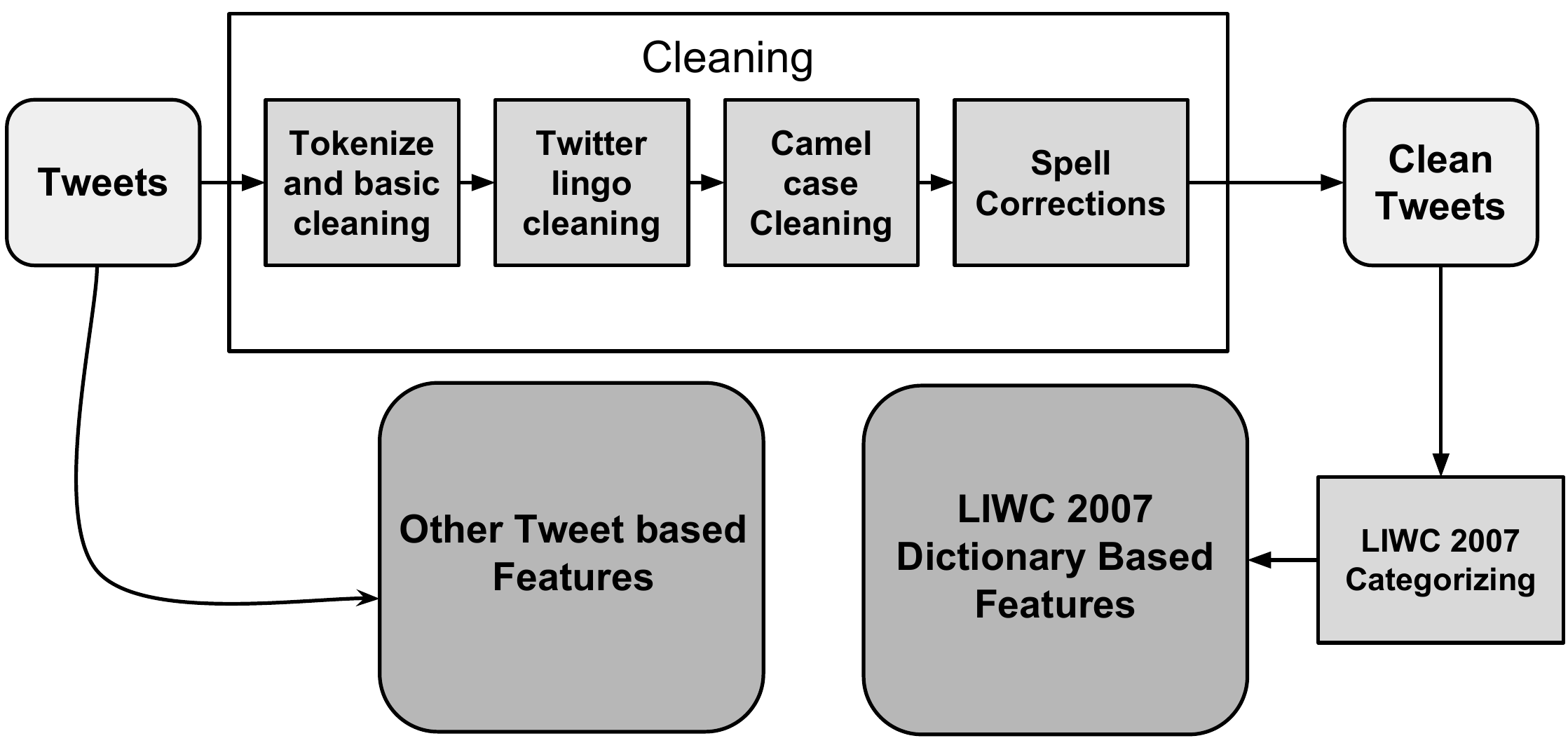}
\caption{High level view of tweet based features extraction}
\label{overflow}
\end{figure}

\section{Features}
Previous works in this area have already explored some general lexicon features based on tweet text for e.g. K-top words used by each class.
In this paper we have not used these features, instead we used LIWC 2007 dictionary for extracting lexicon features. With LIWC based lexicon
features we got almost 3\% gain in accuracy compared to state-of-the-art lexicon features as shown in Section 6.
\\
We will now discuss all the features used in this paper for predicting gender of Twitter user.

\subsection{Tweet behavior based features}
Tweeting behavior can be a good parameter to distinguish between male and female, we have used almost all the features used by state-of-the-art.
\subsubsection{Tweet Frequency}
Let's say number of tweets extracted for a user be \textit{N} (in this paper we used N=1000) and 
chronological difference in days between first and last tweet be \textit{C}, then ``tweet frequency'' is defined as
\begin{equation}
TF = \frac{N}{C}
\end{equation}
According to a study done by \textit{www.beevolve.com} on 36 million twitter users in 2012, it was found that on average female users tend to tweet more often than male counterparts.
\subsubsection{Hashtag Frequency}
Let ${T}_h$ be the total number of hashtags used by a user, then ``hashtag frequency'' is defined as
\begin{equation}
HF = \frac{{T}_h}{C}
\end{equation}
\subsubsection{Average Tweet Length}
``Average Tweet Length'' $TL$ represents the average length
of tweets. Here, $TL$ is defined as
\begin{equation}
     TL= \frac{\sum \limits^{Recent N}_{i=0}Length(tweet_i)}{N}
\end{equation}

\subsubsection{Retweet Frequency}
Let ${T}_r$ be the total number of tweets retweeted by a user, then ``retweet frequency'' is defined as (Rao et al. 2010):
\begin{equation}
     RF= \frac{{T}_r}{C}
\end{equation}

\subsubsection{Followers to following ratio}
The ratio between number of followers and number of friends has been used as a measure
of a user's tendency towards producing vs. consuming information on Twitter (Rao et al. 2010). 

\subsubsection{Celebrity following tendency}
Let ${T}_p$ be the total number of celebrities or popular users followed by a user and ${T}_f$ be the total number , then `celebrity following tendency``
is defined as
\begin{equation}
     CT= \frac{{T}_p}{C}
\end{equation}

\subsection{Linguistic content based features}
Linguistic content information encapsulates the main topics
of interest to the user as well as the user's lexical usage. In a study done by \textit{www.beevolve.com} (2012),
Female users talk more about family and fashion whereas the Twitter male users prefer technology, sports and entrepreneurship.
We explored a wide variety of linguistic content features using LIWC\footnote{Linguistic Inquiry and Word Count (LIWC) is a text analysis software program designed by James W. Pennebaker, Roger J. Booth, and Martha E. Francis. LIWC calculates the degree to which people use different categories of words across a wide array of texts, including emails, speeches, poems, or transcribed daily speech.} 2007 dictionary,
as detailed below.

\subsubsection{Count of \textit{I}, \textit{we}, \textit{you}, \textit{he-she} and \textit{they} words}
According to David Bamman et al. (2014), female Twitter user tend to use more personal diary writing style, where they might use more
\textit{I} and \textit{he-she} references. 
\subsubsection{Parts of Speech}
\textbf{Pronouns} are generally associated with female authors (David Bamman et al., 2014), \textbf{Conjunction} such as \textit{and} were associated with female authors.
Other parts of speech such as articles, determiners or prepositions showed low gender association.
\subsubsection{Emotions and Emoticons}
Emotional terms such as sad, love, glad, etc. are more associated with female authors, while female author show more associativity with emotionally neutral sentences (David Bamman et al., 2014).
\\*
We have also used Emoticons such as :-), <3 etc. and their unicode counterparts. Emoticons tend to have more associativity with female authors.
\subsubsection{Categorical words}
Words related to Health, Money, Achievement, Society, Family etc. can be a good indicator for predicting gender of a twitter users.
As shown in many studies words related to ``Money and'' ``Finance'' are often used by male authors, while words related to ``Family'', ``Kinship'' and ``Society'' are more often
used by female authors on Twitter. Similarly ``Swear'' words are more often used by male authors rather than female authors (David Bamman et al., 2014; \textit{www.beevolve.com}, 2012).
In this papers we have used 50 such categories, each as one feature.
\begin{table}[!htbp]
\centering

\begin{tabular}{|c|c|}
\hline
\textit{\textbf{Categories}}                & \textit{\textbf{Associativity}} \\ \hline
Pronouns                           & Female                                        \\ \hline
Emotion terms                      & Female                                        \\ \hline
Kinship terms                      & Female                                        \\ \hline
CMC words (lol, omg)               & Female                                        \\ \hline
Numbers                            & Male                                         \\ \hline
Technology words                   & Male                                        \\ \hline
Swear words                        & Male                                        \\ \hline
Assent                             & Female                                         \\ \hline
Emoticons                          & Female                                        \\ \hline
Hesitation                         & Female                                        \\ \hline
\end{tabular}
\caption{Some of the categories and their known associativity with gender of user (David Bamman et al., 2014)}
\end{table}
\\
\\
In total we are using 64 different Linguistic content based features. These features proved to give better results than other simple Linguistic content based features
such as \textit{K-top} words. 

\subsection{Popular neighborhood based features}
As we have already defined ``popular neighborhood'' or ``celebrity users'' as those users who are followed by any of the 265 users in our dataset having either (1) more than 10,000 followers or (2) are verified users\footnote{In 
June 2008, Twitter launched a verification program, allowing celebrities, brands, businesses and public figures to get their accounts verified, there are currently approximately 92,000 Verified users on Twitter}.
Popular users (celebrities) use their actual name on Twitter and it is feasible to extract their features from websites like Wikipedia and Freebase. This gives us an opportunity to leverage
principle of homophilic association (in Twitter, a user
can exercise greater selectivity over who she follows than
who follows her) with greater accuracy in extracting neighborhood features.
\\
In the following subsections we would discuss our intuition behind each ``popular neighborhood based feature'' we have chosen and how we extracted those features.
\subsubsection{Age of celebrity}
We have used 5 categories(each as a feature) for a celebrity users age
\begin{description}
\item[$\bullet$] age < 23 years
\item[$\bullet$] 23<age<30
\item[$\bullet$] 30<age<40
\item[$\bullet$] 40<age<50
\item[$\bullet$] 50<age
\end{description}
for e.g. Justin Bieber would be of ``age<23'' category. The basic intuition behind taking celebrity age as feature for predicting gender is,
number of female users under age 30 years on Twitter outnumbered males under age 30, this gap even widens under age 20. Concept of homophilic association
suggests that users of same age category tend to follow each other more. Thus, probability of female following celebrities under age 23 years 
would be more than male. We used Wikipedia to extract age of the celebrity.
Later in our experiment with this feature, we found that age of a celebrity has a weak associativity with the gender of its follower.

\subsubsection{Gender of celebrity}
It has been discussed by Bill Heil et al. (2009), that an average man is almost twice more likely to follow another man than a woman.
We have used ``male'' and ``female'' category of celebrity as feature(for e.g. Sachin Tendulkar would be categorized as male), and ignored gender neutral Twitter users such as Brand and Business accounts.
\begin{table}[h]
\centering

\begin{tabular}{|c|c|c|}
\hline
                                 & \textit{a female}                & \textit{a male} \\ \hline
Female follows                    & 44\%                            & 56\%           \\ \hline
Male follows                      & 35\%                            & 65\%         \\ \hline
\end{tabular}
\caption{Tendency of male and female following other male and female (Bill Heil et al., 2009)}
\end{table}
\\
Since it is almost impossible to find gender of the complete neighborhood of a user, we have used just popular neighborhood. 
We extracted gender of a celebrity using Wikipedia by counting \textit{he, she, her, his, him etc.} references on that celebrities page.
In our experiment we found promising evidence suggesting gender of neighborhood could be helpful in predicting gender of user itself.

\subsubsection{Celebrities ``famous for''}
Work of Lim and Datta (2012) suggests that celebrities represent an interest category, and they leveraged these interests to discover communities who follow specific interests on Twitter.
In this paper we have also used similar categories to find interests of a user using the kind of celebrity they follow. We have used followed following categories each as one feature.
\\
\begin{table}[!htbp]
\centering
\begin{tabular}{|c|}
\hline
 \textbf{``famous for'' Categories of celebrities}
\\ \hline

\begin{tabular}{l l}

$\bullet$ Acting & $\bullet$ Art \\
$\bullet$ Entertainment & $\bullet$ Entrepreneur \\
$\bullet$ Writing & $\bullet$ Music \\
$\bullet$ Politics & $\bullet$ Religious \\
$\bullet$ Science \& Technology & $\bullet$ Security \\
$\bullet$ Social & $\bullet$ Sports \\
$\bullet$ Miscellaneous & \\
\end{tabular}
\\ \hline
\end{tabular}
\end{table}
\\

According to a study conducted by \textit{www.beevolve.com} on 36 million twitter users, it was found that some of the broad categories were more
distinctive then other. And these are of interests could be used predict gender and other attributes of a user.
\\
\begin{figure}[!htbp]
\centering
\includegraphics[width=85mm]{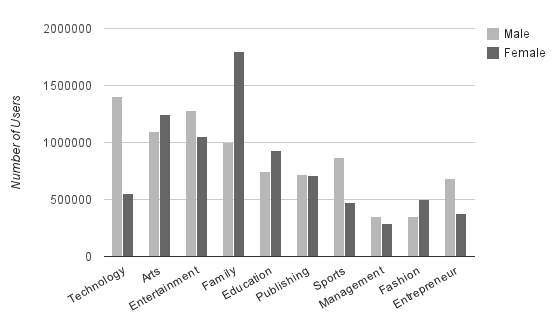}
\caption{Gender distribution by top 10 categories of interest on Twitter (\textit{www.beevolve.com}, 2012)}
\label{overflow}
\end{figure}
\\
\\
\\
\\
\\
\\
\\
\section{Results}
We would evaluate our results in terms of overall accuracy, A 10-fold cross-validation
was done to assess the performance of the SVM-based classifiers at inferring the gender of a Twitter user.
\\*
We will discuss the results in following three stages:
\subsection{Using tweet behavior and linguistic features}
\begin{table}[!htbp]
\centering

\begin{tabular}{|c|c|}
\hline
\textit{\textbf{Configuration}}                & \textit{\textbf{Accuracy}} \\ \hline
LIWC only(with last 200 tweets)    & 77.95\%                                        \\ \hline
LIWC only(with last 1000 tweets)   & 80.43\%                                        \\ \hline
LIWC + Tweet Behavior features     & \textbf{82.26\%}                                        \\ \hline
\end{tabular}
\caption{The overall accuracy while using all the tweet based features which excludes homophily features}
\end{table}

\subsection{Using popular neighborhood features}
\begin{table}[!htbp]
\centering

\begin{tabular}{|c|c|}
\hline
\textit{\textbf{Configuration}}                & \textit{\textbf{Accuracy}} \\ \hline
Age of celebrity		& 56.05\%                                        \\ \hline
Celebrity ``famous for''   	& 67.23\%                                        \\ \hline
Gender of celebrity   	& \textbf{71.57\%}                                        \\ \hline
Combined    	& 66.84\%                                        \\ \hline
\end{tabular}
\caption{The overall accuracy using popular neighborhood features}
\end{table}

\subsection{Using all the features combined}
\begin{table}[!htbp]
\centering
\begin{tabular}{|c|c|c|l|}
\hline
                                 & \textit{True Male}                & \textit{True Female} & \textbf{Precision} \\ \hline
Pred. Male                    & 115                            & 17           & 87.12\%\ \\ \hline
Pred. Female                      & 21                            & 112        & 84.21\% \\ \hline
\textbf{Recall}                      & 84.56\%                            & 86.82\%        & \\ \hline
\end{tabular}
Overall accuracy = \textbf{85.67\%}
\caption{The final confusion matrix and overall accuracy using all the features}
\end{table}

\section{Conclusion and future}

In this paper, we evaluated the extent to which features
present in a Twitter user’s popular neighbors can improve
the inference of attributes possessed by the user herself.
Our results support several noteworthy conclusions, which we discuss here.

\subsection{Usefulness of linguistic features}
Liguistic features such as associativity of a particular category of words towards the gender of a user as suggested by David Bamman et al. (2014), produced satisfactory results.
\\*
\subsection{Popular neighborhood is useful and feature rich}
Pennacchiotti et. al (2011), discussed in their paper that users following more celebrities tend to give better results in their homophily(network) based approach. We verified in this paper that 
popular neighborhood can be leveraged to extract various rich and accurate features using already available knowledge bases.\\
We have observed that gender of a celebrity can be a good feature, and proves the study conducted by Bill Heil et al. (2009), which suggests tendency of males to follow male is twice than following female.\\
Information regarding popular user's domain of work can be useful for determining users interest which in turn can be useful in gender classification.
\\ \\
In this paper we have not used other homophily features such as linguistic features of popular neighborhood, 
it would interesting to see combined results of homophily and celebrity based features used in this paper. We can also explore the possibility of using these features to infer other user attributes such as ``age'' and ``political orientation''

\section{Acknowledgments}
I would like to thank Rahul Tejwani for his help and active participation in various discussions for this work.
I would also like to thank Professor Rohini Srihari and all those who helped me throughout this project.

\bibliographystyle{abbrv}
\bibliography{sigproc}  % sigproc.bib is the name of the Bibliography in this case
\nocite{*}
% ACM needs 'a single self-contained file'!
%\balancecolumns

\end{document}